\documentclass{elsart}

\usepackage{amsmath,amssymb}
\usepackage{cite}
\usepackage{graphicx}
\usepackage{psfrag}

\newcommand{\mathnotation}[2]{\newcommand{#1}{\ensuremath{#2}}}

\newcommand{\eqlabel}[1]{\label{eq:#1}}
\newcommand{\seclabel}[1]{\label{sec:#1}}

\newcommand{\figlabel}[1]{\label{fig:#1}}

\renewcommand{\eqref}[1]{(\ref{eq:#1})}

\newcommand{\secref}[1]{Section~\ref{sec:#1}}

\renewcommand{\l}{\left}			
\renewcommand{\r}{\right}			
\mathnotation{\lPB}{\left\{}			
\mathnotation{\rPB}{\right\}}			
\mathnotation{\com}{\,{\mathchar"213B}\,}	
\mathnotation{\ldef}{\mathrel{\raisebox{.069ex}{:}\!\!=}}
\mathnotation{\rdef}{\mathrel{=\!\!\raisebox{.069ex}{:}}}
\mathnotation{\pd}{\partial}			
\mathnotation{\grad}{\nabla}			
\mathnotation{\Tangent}{T}			

\mathnotation{\LieA}{\mathfrak{g}}		
\mathnotation{\SO}{\mathrm{SO}}			
\mathnotation{\SU}{\mathrm{SU}}			
\mathnotation{\SOthree}{\SO(3)}			
\mathnotation{\SOtwoone}{\SO(2,1)}		
\mathnotation{\sothree}{so(3)}			
\mathnotation{\sotwoone}{so(2,1)}		
\newcommand{\Trans}[1]{#1^T}			
\newcommand{\norm}[1]{\l\|#1\r\|}		
\newcommand{\gradvec}[2]{{\grad_{#1}\,{#2}}}	
\newcommand{\PBrak}[2]{\lPB#1\com#2\rPB}	

\mathnotation{\reals}{{\Rset}}			
\mathnotation{\SD}{{\mathrm{SD}}}		
\mathnotation{\LB}{{\mathrm{Leib}}}		
\mathnotation{\Cas}{C}				
\mathnotation{\Pinv}{P}				
\mathnotation{\angmomm}{\ell}			
\mathnotation{\angmom}{\boldsymbol{\angmomm}}	
\mathnotation{\angvel}{\boldsymbol{\omega}}	
\mathnotation{\advam}{\alpha}			
\mathnotation{\advbm}{\beta}			
\mathnotation{\adva}{\boldsymbol{\advam}}	
\mathnotation{\advb}{\boldsymbol{\advbm}}	
\mathnotation{\momIner}{I}			
\mathnotation{\posa}{{\mathbf{a}}}		
\mathnotation{\posb}{{\mathbf{b}}}		
\mathnotation{\posam}{{a}}			
\mathnotation{\posbm}{{b}}			
\mathnotation{\funcf}{f}			
\mathnotation{\funcg}{g}			
\mathnotation{\Ham}{H}				
\mathnotation{\HamK}{K}				
\mathnotation{\HamP}{V}				
\mathnotation{\twpar}{\varepsilon}		
\mathnotation{\dotangmom}{{\skew{6}\dot\angmom}}
\mathnotation{\dotadva}{{\skew{1}\dot\adva}}	
\mathnotation{\dotadvb}{{\skew{6}\dot\advb}}	
\mathnotation{\dotadvbm}{{\skew{6}\dot\advbm}}	
\mathnotation{\eulphi}{\phi}			
\mathnotation{\eulpsi}{\psi}			
\mathnotation{\eultheta}{\theta}		
\mathnotation{\eulpphi}{{p_\eulphi}}		
\mathnotation{\eulppsi}{{p_\eulpsi}}		
\mathnotation{\eulptheta}{{p_\eultheta}}	
\mathnotation{\Arot}{A}				
\mathnotation{\Lmom}{L}				
\mathnotation{\eul}{{\mathbf{q}}}		
\mathnotation{\pp}{{\mathbf{p}}}		
\mathnotation{\p}{p}				
\mathnotation{\vv}{{\mathbf{v}}}		
\mathnotation{\ww}{{\mathbf{w}}}		
\mathnotation{\vc}{v}				
\mathnotation{\wc}{w}				
\mathnotation{\vm}{\vc}				
\mathnotation{\wm}{\wc}				
\mathnotation{\vvperp}{\vv_{\bot}}
\mathnotation{\vperp}{\vc_{\bot}}
\mathnotation{\vpara}{\vc}
\mathnotation{\Cmvp}{\eta}
\mathnotation{\unvperp}{{\mathbf{\hat e}}_\bot}
\mathnotation{\unvpara}{{\mathbf{\hat e}}_\parallel}
\mathnotation{\unvi}{{\mathbf{\hat e}}_1}
\mathnotation{\unvii}{{\mathbf{\hat e}}_2}
\mathnotation{\unviii}{{\mathbf{\hat e}}_3}

\newcommand{\LP}{Lie--Poisson}
\newcommand{\ie}{i.e.}

\begin{document}

\begin{frontmatter}

\title{The Twisted Top}

\author{Jean-Luc Thiffeault\thanksref{jltemail}}
\address{Department of Applied Physics and Applied Mathematics,\\
Columbia University, New York, U.S.A.}

\author{P. J. Morrison\thanksref{pjmemail}}
\address{Institute for Fusion Studies and Department of Physics, \\
University of Texas at Austin, Austin, Texas, U.S.A.}

\thanks[jltemail]{Electronic mail: jeanluc@mailaps.org} 
\thanks[pjmemail]{Electronic mail: morrison@physics.utexas.edu}

\date{21 March 2001}

\begin{abstract}

We describe a new type of top, the twisted top, obtained by appending a
cocycle to the Lie--Poisson bracket for the charged heavy top, thus breaking
its semidirect product structure.  The twisted top has an integrable case that
corresponds to the Lagrange (symmetric) top.  We give a canonical description
of the twisted top in terms of Euler angles.  We also show by a numerical
calculation of the largest Lyapunov exponent that the Kovalevskaya case of the
twisted top is chaotic.

\end{abstract}
\end{frontmatter}

\section{Introduction}

We present a new top, called the twisted top, obtained by modifying the \LP\
bracket for the charged heavy top.  The charged heavy top, also introduced in
this paper, is a heavy top~\cite{Sudarshan,Vinogradov1977,Guillemin} immersed
in an electric field.  The bracket for the charged heavy top arises from a
semidirect product of~$\SOthree$ and~\hbox{$\reals^3 \times \reals^3$}.  The
twisted top is not a top in the classical sense of a rigid body in a
gravitational field.  Rather, it is a mathematical construction obtained by
using a different Lie group to build the \LP\ bracket for the system.  This
abstract procedure is analogous to the manner in which tops are derived
for~$\SO(N)$~\cite{Ratiu1981}, for~$\SU(N)$ (obtained in Hamiltonian
truncations of the Euler equation~\cite{Fairlie1989,Zeitlin1991}), and for
other groups~\cite{Fairlie1998a,Ueno1998}.  The construction method of the
twisted top is also related to tops obtained by deformations of
algebras~\cite{Tsiganov1999}.

The bracket for the twisted top results from adding a cocycle to the charged
heavy top bracket, so that the structure is no longer semidirect.  Such
brackets are classified in Thiffeault~\cite{Thiffeault1998diss} and Thiffeault
and Morrison~\cite{Thiffeault2000}, and the case we are considering is the
simplest example of a Leibniz
extension~\cite{Parthasarathy1976,Thiffeault1998diss,Thiffeault2000}.  Because
we are interested in how the nontrivial cocycle affects the dynamics of the
system, we use the same Hamiltonian for the twisted top as for the heavy top.
The bracket for the twisted top possesses three Casimir invariants, one of
which differs from that possessed by the charged heavy top.

A most interesting feature of the twisted top is that it retains the
integrability property of the Lagrange top: it is integrable when it has an
axis of symmetry (two moments of inertia are equal), its centre of rotation
lies on the symmetry axis, and the electric field vanishes (or, equivalently,
the top is uncharged).  The conserved quantities are the energy, the angular
momentum along the symmetry axis, and a third invariant which is a
modification of the conserved component of the canonical momentum in the
Lagrange case.

The outline of this paper is as follows.  In~\secref{chtop} we discuss the
charged heavy top and describe its invariants and some of its integrable
cases.  In~\secref{twtop} we introduce the twisted top and its invariants.
We show that it has an integrable case analogous to the Lagrange case of the
heavy top.  We give a canonical description of the twisted top in terms of
Euler angles in~\secref{candescr}.  In canonical coordinates the difference
between the twisted top and the charged heavy top is transferred from the
bracket to the Hamiltonian, and appears as a term that can be interpreted as a
momentum-dependent potential.  Finally, in~\secref{kov} we discuss our results
and show by numerical calculation that the Kovalevskaya case of the twisted
top is not integrable.

\section{The Charged Heavy Top}
\seclabel{chtop}

Consider a heavy, charged top in constant gravitational and electric fields.
The angular momentum vector is denoted by~$\angmom$, the position of the
centre of mass is a vector~$\posa$, and the position of the centre of charge
is~$\posb$.  The direction and strength of the fixed gravitational and
electric forces are given by the vectors~$\adva$ and~$\advb$, respectively.
The frame of reference is the body frame, so that~$\posa$ and~$\posb$ are
constant.  The energy of such a top is
\begin{equation}
	\Ham(\angmom,\adva,\advb)
		= \half\angmom\cdot\angvel
		+ \adva\cdot\posa
		+ \advb\cdot\posb
	\eqlabel{chtopHam}
\end{equation}
where~\hbox{$\angvel \ldef {\momIner^{-1}}\angmom$} is the angular velocity
and~$\momIner$ is the moment of inertia tensor, which can be taken to be
diagonal by an appropriate choice of frame.  We assume that the top's rotation
is slow enough that the magnetic fields set up by the motion of charges is
negligible, and that the top is a perfect insulator, so that the centre of
charge remains fixed within the body.  The charge, like the mass, does not
have to be distributed uniformly, but only the centres of charge and mass
couple to uniform gravitational and electric fields.

The vectors~$\adva$ and~$\advb$, being fixed in space, rotate in the body
frame.  The dynamics of such a configuration can be generated by a \LP\
bracket with a semidirect product structure,
\begin{multline}
	\PBrak{\funcf}{\funcg}_\SD =
		-\angmom\cdot\l(\gradvec{\angmom}{\funcf}
		\times\gradvec{\angmom}{\funcg}\r)
		- \adva\cdot\l(\gradvec{\angmom}{\funcf}
		\times\gradvec{\adva}{\funcg} + \gradvec{\adva}{\funcf}
		\times\gradvec{\angmom}{\funcg}\r)\\
	- \advb\cdot\l(\gradvec{\angmom}{\funcf}
		\times\gradvec{\advb}{\funcg}
		+ \gradvec{\advb}{\funcf}
		\times\gradvec{\angmom}{\funcg}\r),
	\eqlabel{chtopbrak}
\end{multline}
where~$\funcf$ and~$\funcg$ are functions of~\hbox{$(\angmom,\adva,\advb)$},
and~$\gradvec{}{}$ is a gradient with respect to its subscript.
Equation~\eqref{chtopbrak} is a simple extension of the bracket for the heavy
top, which also has a semidirect product
structure~\cite{Vinogradov1977,Guillemin,Marsden1984,Holmes1983}
(without~$\adva$).  The Casimir invariants of Eq.~\eqref{chtopbrak} are
\begin{equation*}
	\Cas_1 = \norm{\adva}^2, \quad
	\Cas_2 = \adva\cdot\advb, \quad
	\Cas_3 = \norm{\advb}^2.
\end{equation*}
The invariant~$\Cas_2$ says that the angle between~$\adva$ and~$\advb$ is
constant, because by~$\Cas_1$ and~$\Cas_3$ their length is conserved.
Therefore, the two vectors~$\adva$ and~$\advb$ fully describe the orientation
of the rigid body, and there is a one-to-one mapping between~$\adva$
and~$\advb$ and Euler angles.  The phase space of the motion is
thus~$\SOthree\times\Rset^3 \cong \Tangent^*\SOthree$.  This is the same phase
space as in the unreduced (canonical) system~\cite{Thiffeault1998}.

For the case with~$\momIner_1=\momIner_2$, \hbox{$\posa =
\Trans{(0,0,\posam_3)}$}, and~\hbox{$\posb = 0$}, the charged heavy top
reduces to the Lagrange top, also called the heavy symmetric top, and so is
integrable (see for example Audin~\cite{Audin}).  The invariants are the
energy~$\Ham$,~$\angmomm_3$, and~\hbox{$\angmom\cdot\adva$}.  (There is also
an integrable case with~\hbox{$\posb = \Trans{(0,0,\posbm_3)}$} and
\hbox{$(\adva\times\advb)\cdot\posb = 0$}, \ie, where the forces and
the centre of charge are coplanar.)

\section{The Twisted Top}
\seclabel{twtop}

In Thiffeault~\cite{Thiffeault1998diss} and Thiffeault and
Morrison~\cite{Thiffeault2000}, it is shown that the only bracket extension of
two field variables (such as~$\angmom$ and~$\adva$) is of the semidirect
product type.  To obtain an extension that is not semidirect, one requires at
least three variables, which we take to be the same variables as for the
charged heavy top.  The simplest non-semidirect extension is then the Leibniz
bracket
\begin{equation}
	\PBrak{\funcf}{\funcg}_\LB = \PBrak{\funcf}{\funcg}_\SD
		- \twpar\,\advb\cdot\l(\gradvec{\adva}{\funcf}
		\times\gradvec{\adva}{\funcg}\r),
	\eqlabel{twtopbrak}
\end{equation}
where~$\twpar$ is a parameter measuring the deviation from a semidirect
bracket and is not necessarily small.  Using the same
Hamiltonian~\eqref{chtopHam} as for the charged heavy top in the
bracket~\eqref{twtopbrak}, we obtain the equations
\begin{alignat}{2}
	\dotangmom &= \PBrak{\angmom}{\Ham}_\LB &= \angmom\times\angvel
		&+ \adva\times\posa + \advb\times\posb,
		\eqlabel{dotangmom}\\
	\dotadva &= \PBrak{\adva}{\Ham}_\LB &= \adva\times\angvel &+
		\twpar\,\advb\times\posa,
		\eqlabel{dotadva}\\
	\dotadvb &= \PBrak{\advb}{\Ham}_\LB &= \advb\times\angvel&.
		\eqlabel{dotadvb}
\end{alignat}
These are the equations for the twisted top.  The term proportional
to~$\twpar$ in the~$\dotadva$ equation adds a ``twist'' which means
that~$\adva$ does not simply rotate rigidly (though~$\advb$ still does).  This
is reflected in the Casimir invariants, which are now
\begin{equation}
	\Cas_1=\norm{\adva}^2 + 2\twpar\,\angmom\cdot\advb, \quad
	\Cas_2=\adva\cdot\advb, \quad
	\Cas_3=\norm{\advb}^2.
	\eqlabel{twcas}
\end{equation}
Since the length of~$\adva$ is no longer
preserved, the invariant~$\Cas_2$ does not imply that the angle
between~$\adva$ and~$\advb$ is constant.  However, the length of the
\emph{projection} of~$\adva$ onto~$\advb$ is preserved.

For a positive-definite moment of inertia tensor, the energy surfaces of the
twisted top are bounded, as can be seen from the following argument.  First
note that the components of~$\angmom$ cannot diverge without~$\adva$
or~$\advb$ also diverging, since~$\angmom$ enters the Hamiltonian in a
positive-definite quadratic form.  From the invariant~$\Cas_3$, the components
of $\advb$ are finite.  To have unbounded surfaces, and still
conserve~$\Cas_1$, both~$\norm{\angmom}$ and~$\norm{\adva}$ must go to
infinity, with~\hbox{$\norm{\adva}^2\sim-2\twpar\,\angmom\cdot\advb$}.  But
with this functional relation it is not possible to
have~$\norm\angmom\rightarrow\infty$ whilst preserving the Hamiltonian~$\Ham$,
since its kinetic part is proportional to~$\norm{\angmom}^2$ and its potential
part to~\hbox{$\norm{\adva}\sim\norm{\angmom}^{1/2}$}, precluding any balance.
We conclude that the energy surfaces are bounded.  This will be important
in~\secref{kov} where we try to demonstrate chaotic behaviour by computing the
largest Lyapunov exponent.

The twisted top also has an integrable Lagrange case.  It is obtained, as for
the charged heavy top, by letting~\hbox{$\momIner_1=\momIner_2$}, \hbox{$\posa
= \Trans{(0,0,\posam_3)}$}, and~\hbox{$\posb = 0$}.  The energy~$\Ham$ and the
third component of the angular momentum~$\angmomm_3$ are still conserved,
whereas the third invariant becomes
\begin{equation}
	\Pinv = \angmom\cdot\adva +
	\twpar\,\momIner_1\posam_3\,\advbm_3\,.
	\eqlabel{Pinv}
\end{equation}
We call this integrable case the twisted Lagrange top.  We can verify
that~$\Pinv$ is conserved directly from the equations of
motion~\eqref{dotangmom}--\eqref{dotadvb},
\begin{align*}
	\dot\Pinv &=
		\dotangmom\cdot\adva
		+ \angmom\cdot\dotadva
		+ \twpar\,\momIner_1\posam_3\dotadvbm_3\\
	&= (\angmom\times\angvel)\cdot\adva
		+ \angmom\cdot(\adva\times\angvel
		+ \twpar\,\advb\times\posa)
		+ \twpar\,\momIner_1\,\posam_3\,(\advb\times\angvel)_3\\
	&= \twpar\,\posa\cdot(\angmom\times\advb)
		+ \twpar\,\posa\cdot(\advb\times\momIner_1\,\angvel) = 0,
\end{align*}
where we equated~$\momIner_1\,\angvel$ to~$\angmom$ in the last triple product
because only the first two components of~$\angvel$ are involved,
and~$\momIner_1=\momIner_2$.  It is straightforward to verify that the
invariants~$\{\Ham,\angmomm_3,\Pinv\}$ are in involution, \ie, they commute
with respect to the bracket~\eqref{twtopbrak}---a necessary condition for
integrability.  The commutativity of the invariants carries over to the
canonical variables of~\secref{candescr}.

\section{Canonical Description}
\seclabel{candescr}

Since the twisted top is a Hamiltonian system, there exists a coordinate
transformation on the symplectice leaves (the constraint surfaces described by
the Casimirs) that makes the system canonical.  We now proceed to find such a
coordinate transformation, in a manner analogous to the reduction of the rigid
body and the heavy top~\cite{Holmes1983,Morrison1998}.  The transformation we
describe will be from the three Euler angles~\hbox{$\eul =
\Trans{(\eulphi,\eulpsi,\eultheta)}$} and their corresponding canonical
momenta~$\pp = \Trans{(\eulpphi,\eulppsi,\eulptheta)}$ (6 coordinates) to the
vectors~$(\angmom,\adva,\advb)$ (9 coordinates, 3 Casimirs).  We show that the
transformation is invertible on the symplectic leaves, so that it can be used
to canonize the system.

Following the heavy top reduction~\cite{Holmes1983}, since the vector~$\advb$
rotates rigidly (length conserved), it is fixed in the space frame, and we
write
\begin{equation}
	\advb = \Arot(\eul)\,\ww,
	\eqlabel{advbtrans}
\end{equation}
where the rotation matrix~$\Arot$ is
\begin{equation*}
	\begin{pmatrix}
	\cos\eulpsi\cos\eulphi - \cos\eultheta\sin\eulphi\sin\eulpsi &
	\cos\eulpsi\sin\eulphi + \cos\eultheta\cos\eulphi\sin\eulpsi &
	\sin\eulpsi\sin\eultheta \\
	-\sin\eulpsi\cos\eulphi - \cos\eultheta\sin\eulphi\cos\eulpsi &
	-\sin\eulpsi\sin\eulphi + \cos\eultheta\cos\eulphi\cos\eulpsi &
	\cos\eulpsi\sin\eultheta \\
	\sin\eultheta\sin\eulphi & -\sin\eultheta\cos\eulphi & \cos\eultheta
	\end{pmatrix}.
\end{equation*}
The matrix~$\Arot$ transforms vectors from the space frame to the body frame
(we are following the convention of Goldstein~\cite[p.~147]{Goldstein} for the
definition of~$\Arot$.)  The vector~$\ww$ is constant and fixed in space.
Since rotations preserve lengths, we have~\hbox{$\Cas_3 = \norm{\advb}^2 =
\wm^2$}, where~\hbox{$\wm = \norm{\ww} \ge 0$}.

For the angular momentum, we take
\begin{equation}
	\angmom = \Lmom(\eul)\,\pp,
	\eqlabel{angmomtrans}
\end{equation}
where~$\Lmom$ is more concisely defined via its inverse,
\begin{equation*}
	\Lmom^{-1} \ldef
	\begin{pmatrix}
	\sin\eultheta\sin\eulpsi\ & \sin\eultheta\cos\eulpsi\ &
		\cos\eultheta\ \\
	0 & 0 & 1 \\
	\cos\eulpsi & -\sin\eulpsi & 0
	\end{pmatrix}.
\end{equation*}
This is the usual transformation one makes when reducing the rigid
body~\cite[p.~499]{Morrison1998}, where~$\Lmom$
maps~$\Tangent^*_{\eul}\,\SOthree$ to~$\LieA^*$.

Finally, for~$\adva$ we try the form
\begin{equation}
	\adva =
	\Arot(\eul)\,\vv(\eul,\pp),
	\eqlabel{advavvdef}
\end{equation}
where we have allowed~$\vv$ to depend on the Euler angles and canonical
momenta in an effort to conserve the Casimirs~$\Cas_1$ and~$\Cas_2$.  We then
have
\[
	\Cas_2 = \adva\cdot\advb
	= \vv\cdot\Trans{\Arot}\Arot\,\ww
	= \vv\cdot\ww,
\]
where we used the orthogonality of~$\Arot$.  Decomposing~$\vv$ into a
part~$\vperp\,\unvperp$ perpendicular to~$\ww$ and a part~$\vpara\,\unvpara$
parallel to~$\ww$, we obtain~\hbox{$\Cas_2 = \vpara\,\wm$}.  Since~$\wm$ is
constant, we require~$\vpara$ to also be constant.

The norm of~$\adva$ is
\begin{equation*}
	\norm{\adva}^2 = \vv\cdot\Trans{\Arot}\Arot\,\vv
	= \vperp^2 + \vpara^2 = \Cas_1 - 2\twpar\,\angmom\cdot\advb,
\end{equation*}
where we used the definition~\eqref{twcas} of~$\Cas_1$.  We solve this
for~$\vperp^2$,
\begin{equation}
	\vperp^2 = \Cas_1 - \vpara^2 - 2\twpar\,\angmom\cdot\advb
	= \norm{\adva}^2 - \frac{(\adva\cdot\advb)^2}{\advb^2} \ge 0,
	\eqlabel{vperp2}
\end{equation}
with~\hbox{$\vperp^2 = 0$} if and only if~$\adva$ and~$\advb$ are collinear.
For convenience, define the constant
\begin{equation*}
	\Cmvp \ldef \Cas_1 - \vpara^2
	= \norm{\adva}^2 - \frac{(\adva\cdot\advb)^2}{\advb^2}
	+ 2\twpar\,\angmom\cdot\advb,
\end{equation*}
which we will use from now on instead of~$\Cas_1$.  Then Eq.~\eqref{advavvdef} becomes
\begin{equation}
	\adva =
	\Arot\l[\vpara\,\unvpara
	+ \sqrt{\Cmvp - 2\twpar\,\angmom\cdot\advb}\,\unvperp\r]
	\eqlabel{advatrans}
\end{equation}
Note that the vectors~$\adva$ and~$\advb$ are collinear
(\hbox{$\adva\times\advb=0$}) if and only if \hbox{$\Cmvp =
2\twpar\,\angmom\cdot\advb$}.  Assume that they are initially not collinear
(\hbox{$\Cmvp \ne 2\twpar\,\angmom\cdot\advb$}).  The time evolution
of~$2\twpar\,\angmom\cdot\advb$ is obtained from~\eqref{dotadva} and the
conservation of~$\Cas_1$, yielding
\begin{equation}
	2\twpar\,\frac{d}{dt}(\angmom\cdot\advb) = -\frac{d}{dt}\norm{\adva}^2
	= -2\twpar\,\posa\cdot(\adva\times\advb).
	\eqlabel{ldotbetaev}
\end{equation}
If~\hbox{$\Cmvp = 2\twpar\,\angmom\cdot\advb$} initially, then it remains so
for all times, because then the right-hand side of~\eqref{ldotbetaev}
vanishes.  Conversely, if~\hbox{$\Cmvp \ne 2\twpar\,\angmom\cdot\advb$}
initially, then the two vectors~$\adva$ and~$\advb$ are never collinear.

This is crucial because it tells us that we can always invert
Eqs.~\eqref{advbtrans} and~\eqref{advatrans}
for~$(\eulphi,\eulpsi,\eultheta)$, as long as~$\adva$ and~$\advb$ are not
initially collinear.  The inversion is done as follows: take~$\ww$ as the
spatial~$z$-axis.  Then~$\advb$ is~$z'$, the transformed~$z$-axis, which
allows us to determine~$\eulpsi$ and~$\eultheta$, but not~$\eulphi$ since it
represents a rotation about the~$z$-axis.  We then use~$\vvperp$ to define
the~$x$-axis, which allows us to find~$\eulphi$ from~$\adva$ (as long
as~\hbox{$\vvperp\ne0$}, but we showed that it is sufficient to require this
initially).  But since the~$2\twpar\,\angmom\cdot\advb$ term only affects the
magnitude of~$\vvperp$, not its orientation, we conclude that the Euler angles
are only a function of~$\adva$ and~$\advb$, not of~$\angmom$.  We can then go
back and solve~\eqref{angmomtrans} for the canonical momenta.
(Provided~\hbox{$\det\Lmom^{-1}=\sin\eultheta\ne 0$}, the coordinate
singularity inherent to Euler angles.  This singularity can be avoided by
``inflating'' the phase space~\cite{Dullin1996}.)

It is straightforward to show that in the
coordinates~\hbox{$(\eul,\pp)$}
the bracket~\eqref{twtopbrak} does indeed have the canonical form.

In canonical coordinates, the `potential' part of the
Hamiltonian~\eqref{chtopHam} becomes
\begin{equation}
\begin{split}
	\HamP(\adva,\advb) &= \adva\cdot\posa + \advb\cdot\posb\\
	&= (\vpara\,\posa + \wm\,\posb)\cdot\Arot\,\unvpara
	+ \sqrt{\Cmvp - 2\twpar\,\angmom\cdot\advb}\,
	\posa\cdot\Arot\,\unvperp
	\eqlabel{twpot}
\end{split}
\end{equation}
with
\begin{equation*}
	\angmom\cdot\advb = \pp\cdot
	\Trans{\Lmom}(\eul)\,\Arot(\eul)\,\ww.
\end{equation*}
The matrix~$\Trans{\Lmom}\Arot$ is
\begin{equation*}
	\Trans\Lmom\Arot =
	\begin{pmatrix}
	-\cot\eultheta\sin\eulphi & \cot\eultheta\cos\eulphi & 1 \\
	\csc\eultheta\sin\eulphi & -\csc\eultheta\cos\eulphi & 0 \\
	\cos\eulphi & \sin\eulphi & 0
	\end{pmatrix}.
\end{equation*}

The integrable case of the twisted top
has~\hbox{$\momIner_1=\momIner_2$},~\hbox{$\posa =
\Trans{(0,0,\posam_3)}$},~$\posb=0$, for which the kinetic energy is
independent of~$\eulphi$ and~$\eulpsi$, and the potential becomes
\begin{equation*}
	\HamP(\eulphi,\eultheta,\pp)
	= \posam_3(\vpara\,\unviii\cdot\Arot\,\unvpara
	+ \sqrt{\Cmvp - 2\twpar\,\angmom\cdot\advb}\,\,
	\unviii\cdot\Arot\,\unvperp).
\end{equation*}
Note that both~$\Trans{\Lmom}\Arot$ and~$\unviii\cdot\Arot =
(\sin\eultheta\sin\eulphi,{-\sin\eultheta\cos\eulphi},\cos\eultheta)$ are
independent of~$\eulpsi$, so that in the twisted Lagrange top case~$\eulpsi$
is cyclic ($\eulppsi$ conserved).

A particularly simple choice is~$\unvpara=\unviii$,~$\unvperp=\unvi$,~$\wm=1$,
for which the potential is
\begin{equation*}
	\HamP(\eulphi,\eultheta,\pp)
	= \posam_3(\vpara\cos\eultheta
	+ \sqrt{\Cmvp - 2\twpar\,\eulpphi}\,\sin\eultheta\sin\eulphi).
\end{equation*}
Though simple, this description does not reduce nicely to the Lagrange top
when~\hbox{$\twpar\rightarrow 0$}.  The
choice~$\unvperp=\unviii$,~$\unvpara=\unvi$,~$\wm=1$, which gives
\begin{equation*}
	\HamP(\eulphi,\eultheta,\pp)
	= \posam_3(\vpara\,\sin\eultheta\sin\eulphi
	+ \sqrt{\Cmvp - 2\twpar\,\angmom\cdot\advb}\,\cos\eultheta),
\end{equation*}
with
\begin{equation*}
	\angmom\cdot\advb = -\eulpphi\cot\eultheta\sin\eulphi
		+ \eulppsi\csc\eultheta\sin\eulphi + \eulptheta\cos\eulphi.
\end{equation*}
has the Lagrange top form when~\hbox{$\vpara=\twpar=0$}, at the cost of a
more complicated expression for~\hbox{$\angmom\cdot\advb$}.

\section{Discussion}
\seclabel{kov}

We have introduced a simple generalisation of the heavy top by giving it
charge and placing it in a constant electric field.  By deforming the bracket
of this charged heavy top, we have obtained a new top that we call twisted.
We have found that the twisted top possesses an integrable case analogous to
the Lagrange top.  It is then natural to ask if such a deformation always
preserves integrability.  An affirmative answer would be very surprising,
considering the delicate nature of integrable systems, and indeed it does not
seem to be so for the system at hand.

We investigate this by looking at the Kovalevskaya case of the twisted top.
For the uncharged heavy top ($\posb=0$, $\twpar=0$), the Kovalevskaya case
involves setting~$\momIner_1 = \momIner_2$,~$\momIner_3 = 2\momIner_1$,
and~$\posam_3=0$.  This top is integrable~\cite{Kowalewski1889,Audin}.  The
analogous case for the twisted top, as for the Lagrange top, simply involves a
change of bracket by setting~$\twpar\ne 0$.  Specifically, we
choose~$\twpar=1$ and~$\posa=\Trans{(-1,0,0)}$.

Figure~\ref{fig:twtop_lyap} shows a plot of the instantaneous largest Lyapunov
exponent of the twisted Kovalevskaya top, after averaging over~$20,000$ random
initial conditions integrated numerically (dotted line).  The solid line is a
least-squares fit to help determine the Lyapunov exponent to greater accuracy,
using an asymptotic form of the averaged exponent~\cite{Tang1996}.  The
Lyapunov exponent is~\hbox{$\lambda \simeq 0.122$}, suggesting that the
twisted Kovalevskaya top is chaotic.  For comparison, the dashed line shows
the same calculation for the untwisted (ordinary) Kovalevskaya top, which
shows the Lyapunov exponent going to zero.  Thus, the twisted case appears to
be chaotic, whilst the untwisted case is not.  We conclude that integrability
does not always survive deformation, contrary to the Lagrange case.  Note that
we can infer chaos from a positive Lyapunov exponent because we showed
in~\secref{twtop} that the motion takes place in a bounded region of phase
space~\cite{Eckmann1985}.  The presence of chaos does not rule out the
existence of an integrable case with parameters ``close'' to the Kovalevskaya
values (in the sense of differing only by terms involving~$\twpar$).  We have
not found such a case.  Another approach would be to use a
Kovalevskaya--Painlev\'{e} analysis to determine other integrable limits of
the twisted top.

A rigorous demonstration that the twisted Kovalevskaya top is chaotic could in
principle be achieved using a Melnikov analysis, as was done by Holmes and
Marsden~\cite{Holmes1983} for the heavy top.  The present case is less
straightforward because of the complicated form of the homoclinic orbits.

It would of course be of great value to find a physical realisation of the
twisted top.  One could use either the noncanonical picture,
Eqs~\eqref{dotangmom}--\eqref{dotadvb}, or the canonical picture, given by the
standard free rigid body Hamiltonian~\cite{Holmes1983} with Eq.~\eqref{twpot}
for a potential.  Regardless of the physical interpretation, the twisted top
remains an object worthy of study in its own right, because of its interesting
integrable case and peculiar geometry.  It would be worthwhile to carry out a
topological classification of the bifurcations of the phases space of twisted
Lagrange top, as was done by Dullin \etal~\cite{Dullin1994} for the
Kovalevskaya top.  This is complicated by the need to find a good surface to
make Poincar\'{e} sections in the canonical coordinate space.

\begin{ack}

The authors thank Holger R. Dullin and Tom Yudichak for their comments and
encouragement.  This work was supported by the U.S. Department of Energy under
contract No.~DE-FG03-96ER-54346.

\end{ack}


\clearpage

\begin{figure}
\psfrag{t}{$t$}
\psfrag{L}{$\langle\lambda\rangle$}
\centerline{\includegraphics[width=\textwidth]{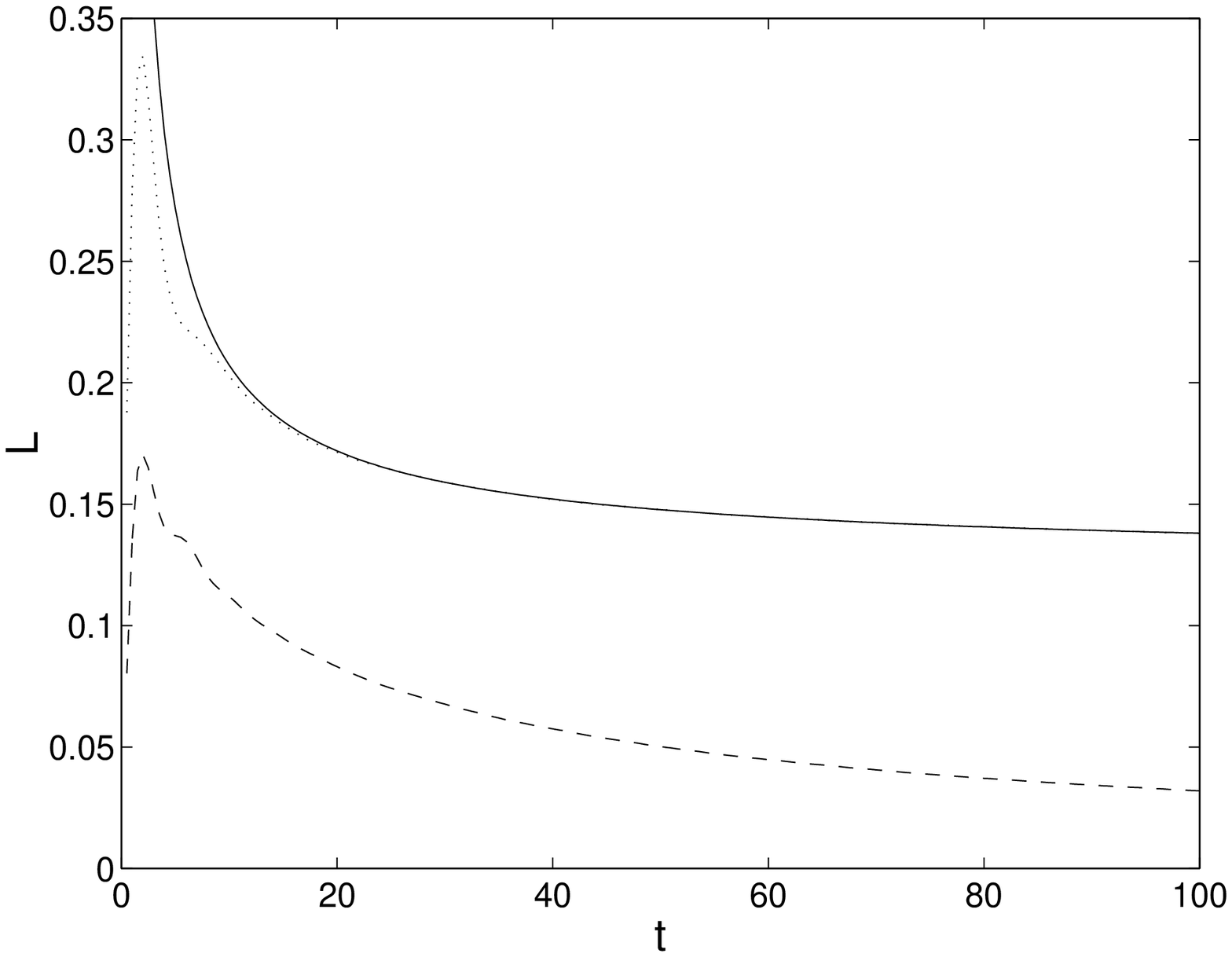}}
\caption{Lyapunov exponent for the twisted top, averaged over initial
conditions.  The dotted line is for the Kovalevskaya top and the solid line is
the function~$0.497/t + 0.113/\sqrt{t} + 0.122$, obtained by a least-squares
fit and yielding the value~\hbox{$\lambda \simeq 0.122$} (See
Ref.~\cite{Tang1996}).  For comparison, the dashed line is the averaged
Lyapunov exponent for the ordinary Kovalevskaya top.}
\figlabel{twtop_lyap}
\end{figure}

\end{document}